\documentclass[12pt]{article}
\usepackage{cite,graphicx,amsmath,amssymb}

\newcommand{\be}{\begin{equation}}  
\newcommand{\ee}{\end{equation}}  
\newcommand{\bea}{\begin{eqnarray}}  
\newcommand{\eea}{\end{eqnarray}}  
\newcommand{\UV}{{\rm UV}}
\newcommand{\IR}{{\rm IR}}
\renewcommand{\Re}{\operatorname{Re}}
\renewcommand{\Im}{\operatorname{Im}}

\begin{document}

\thispagestyle{empty}
\vspace*{.2cm}
\noindent
HD-THEP-05-23 \hfill 14 October 2005
\\
\noindent
HUTP-05-A0047

\vspace*{1.0cm}

\begin{center}
{\Large\bf The Throat as a Randall-Sundrum Model\\[.7cm]
with Goldberger-Wise Stabilization}
\\[1.5cm]
{\large F.~Br\"ummer$\,^a$, A.~Hebecker$\,^a$, and 
E.~Trincherini$\,^{a,\,b}$}\\[.5cm]
{\it ${}^a$ Institut f\"ur Theoretische Physik, Universit\"at Heidelberg,
Philosophenweg 16 und 19, D-69120 Heidelberg, Germany}
\\[.3cm]
{\it ${}^b$ Jefferson Physical Laboratory, Harvard University, 17 Oxford 
Street,\\ Cambridge, Massachusetts 02138, U.S.A.}
\\[.4cm]
{\small\tt (\,f.bruemmer, a.hebecker@thphys.uni-heidelberg.de,} {\small and}
{\small\tt trincher@schwinger.harvard.edu\,)}
\\[1.0cm]

{\bf Abstract}
\end{center} 
An interesting feature of type IIB flux compactifications is the natural 
presence of strongly warped regions or `throats'. These regions allow for a 
5d Randall-Sundrum model interpretation with a large hierarchy between the 
UV and IR brane. We show that, in the 5d description, the flux stabilization 
of this hierarchy (or, equivalently, of the brane-to-brane distance) can be 
understood as an implementation of the Goldberger-Wise mechanism. This 
mechanism relies on the non-trivial bulk profile of the so-called 
Goldberger-Wise scalar, which in addition has fixed expectation values at the 
boundaries and thereby stabilizes the size of the 5d interval. The 
Goldberger-Wise scalar is realized microscopically by the continuously varying 
flux of the Neveu-Schwarz 2-form potential $B_2$ on the $S^2$ cycle in the 
throat. Its back-reaction on the 5d geometry leads to a significant departure 
from a pure AdS$_5$ background. We also find that, for a wide range of 
parameters, the universal K\"ahler modulus of the 10d compactification plays 
the role of a UV-brane field in the equivalent 5d model. It governs the size 
of a large 4d curvature term localized at the UV brane. We hope that our 
simple 5d description of the stabilized throat will be useful in various 
phenomenological and cosmological applications and that refined versions of 
this construction will be able to account for all relevant details of the 
10d model.

\newpage
\section{Introduction}
Flux compactifications of type IIB string theory, or rather its low-energy 
limit type IIB supergravity, provide an attractive path towards potentially 
realistic 4d models of particle physics. Remarkably, by turning on fluxes in 
the supergravity background it is possible to stabilize all complex structure 
moduli and the dilaton in a controlled manner (see~\cite{Polchinski:1995sm, 
Giddings:2001yu} for some of the original papers and~\cite{Frey:2003tf} for 
recent reviews). In addition, in flat 4d compactifications of this type, 
spacetime is in general the warped product (rather than the direct product) 
of Minkowski space and the internal manifold. This can naturally induce a 
large hierarchy of scales.

In the work of Giddings, Kachru and Polchinski~\cite{Giddings:2001yu} the 
construction of such warped supergravity compactifications with fluxes 
and large hierarchies has been analyzed in detail (see also~\cite{ 
DeWolfe:2002nn,Giddings:2005ff}). Spacetime is taken to be the warped product 
of Minkowski space with a compact Calabi-Yau 3-fold. To understand the part 
of the construction relevant to the hierarchy, consider first the unwarped 
case and a 3-fold which has a conifold singularity. This means that the 
neighborhood of a certain point has the geometry of a real cone over the 
compact 5d Einstein manifold $T^{1,1}$ (see e.g.~\cite{Klebanov:1998hh}). 
By smoothing this singularity and turning on fluxes on certain cycles of the 
internal manifold, this region is deformed into a `throat', i.e. a 
strongly warped region of 
spacetime whose geometry is approximately AdS$_5\times T^{1,1}$. The radial 
coordinate of the cone becomes the radial coordinate of AdS; the throat 
connects the remainder of the compact manifold at one end with the blown-up 
tip of the cone at the other end~\cite{Klebanov:1999rd,Klebanov:2000nc, 
Klebanov:2000hb}. The dilaton and the single complex structure modulus of the 
model are stabilized, and for a suitable choice of fluxes, a large hierarchy 
between the two ends is generated~\cite{Giddings:2001yu}. At this level of 
the construction, the universal K\"ahler modulus governing the size of the 
compact space remains massless. It can then be stabilized by nonperturbative 
effects~\cite{Kachru:2003aw,Burgess:2003ic} (see, however,~\cite{Choi:2005ge 
}), by the interplay of perturbative and nonperturbative physics~\cite{ 
Balasubramanian:2004uy}, or even in an entirely perturbative fashion~\cite{ 
Saltman:2004sn,vonGersdorff:2005bf}.

The close relation between type IIB throat geometries and the warped 5d 
models of Randall and Sundrum~\cite{Randall:1999ee,Randall:1999vf} was 
emphasized early on~\cite{Verlinde:1999fy,Chan:2000ms}. Compactifications 
with an infinite AdS$_5\times S^5$ region can be identified precisely with 
the Randall-Sundrum II model. By contrast, the 5d geometry of the stabilized 
$T^{1,1}$ throat relevant here deviates from AdS$_5$ and is only similar to 
the Randall Sundrum I model. The role of the Randall-Sundrum `UV brane' is 
played by the compact manifold, and the `IR brane' is the bottom of the 
throat.\footnote{
We 
alert the reader that we will be using the term `brane' with two different 
meanings, both for string-theoretic D-branes and for the 4d boundaries of
a slice of AdS$_5$ space as is common in field theory model building with
extra dimensions. It should always be clear from the context what kind 
of `brane' we are referring to.
}
The stabilization of the brane-to-brane distance requires that the homogeneity 
of pure AdS space be broken. In the present case, this is realized by the 
deviation from AdS$_5$ geometry observed in the $T^{1,1}$ throat.

It is now natural to inquire about the details of the effective field theory 
describing the 5d physics and, in particular, the stabilization 
mechanism. We consider the understanding of the 5d dynamics to be important 
in view of the wide range of phenomenological applications that such 
constructions might have. We recall the immense amount of cosmological and 
particle physics model building triggered by the Randall-Sundrum proposals
(see~\cite{rspheno} for a selective list of papers and reviews). More 
recently, many interesting applications of the type IIB flux-induced 
hierarchies have emerged (see e.g.~\cite{DeWolfe:2002nn,Giddings:2005ff, 
Kachru:2003aw,throatinf,smallnumbers,Cascales:2005rj,Cascales:2003wn}). In 
particular, it is conceivable that the natural generation of hierarchies by 
fluxes affects the relation of particle phenomenology to the string landscape
and its statistics~\cite{Denef:2004ze}. One of the meeting points of these 
efforts is the 5d effective theory, which, we believe, should therefore be 
understood beyond the leading (pure AdS) approximation. Such an effective 
field theory will be useful independently of whether the Standard Model is 
eventually found to reside on the IR or the UV brane.

In the present paper, we derive some of the essential properties of the 
effective Randall-Sundrum-type model within the following approach: Starting 
from the well-known 10d solution, we identify the coordinate measuring 
physical distances along the throat in 5d units. The fundamental dynamical 
quantity varying along this coordinate is the flux of the Neveu-Schwarz 
2-form potential $B_2$ on the $S^2$ cycle of the $T^{1,1}$ (or, equivalently, 
the 5-form flux on the $T^{1,1}$, or the radius of the $T^{1,1}$). This 
information is encoded in a 5d scalar field $H$ with a well-defined 5d 
profile, which is accompanied by a corresponding profile of the 5d curvature. 
It turns out that, within the 5d model, such a background is easily 
realized by postulating an appropriate 5d potential for $H$. Thus, we find 
ourselves precisely in the setting of the  Goldberger-Wise stabilization 
mechanism~\cite{Goldberger:1999uk} (see also~\cite{gwback}), with a 
slowly varying Goldberger-Wise scalar $H$. (This can be seen as a simplified 
description of one of the multi-scalar solutions of~\cite{Klebanov:2000nc}.) 
The back reaction of the varying scalar potential on the metric deforms 
the AdS space, and by fixing the values of $H$ on the boundaries (introducing 
very steep brane potentials) the length of the throat is also fixed. 
Furthermore, it turns out that, within the validity range of 
the 5d model, the unfixed universal K\"ahler modulus corresponds to a UV brane 
field -- the radius of the 6d compact manifold at the UV end of the throat. 
Thus, we are able to write down an effective 5d action (with appropriate 
brane contributions) incorporating $H$, the universal K\"ahler modulus $\rho$ 
and, of course, the 5d radion governing the length of the 5d compactification 
interval. 

The paper is organized as follows: In Sect.~\ref{warpsugra}, we briefly
review, following Refs.~\cite{Klebanov:1998hh,Klebanov:1999rd,Klebanov:2000nc, 
Klebanov:2000hb} and~\cite{Giddings:2001yu}, the 10d solution for the 
conifold throat to the extent that it will be needed later on. We include a 
simplified derivation of the radial variation of the 5-form flux in the 
throat since the proper 5d description of this effect is one of the main 
objectives pursued in the rest of the paper. We also emphasize that, 
generically, a purely conical region is present between compact space and 
throat. 

In Sect.~\ref{5d}, we turn to the effective 5d model. We relate the 
5d coordinate measuring the invariant distance in the 5d Einstein frame to 
the radial coordinate of the conifold throat. Having identified the 
geometry and the profile of the 5-form flux in terms of this coordinate, we 
are able to construct the corresponding Goldberger-Wise-type stabilization 
model (the potential being $V(H)\sim H^{-8/3}$) and to relate the 
Goldberger-Wise scalar to the 10d quantities varying along the throat.

Section \ref{kaehlerbrane} deals with the 5d equivalent of the universal
K\"ahler modulus. According to~\cite{Giddings:2005ff}, where the explicit 
dependence of warp factor and metric on this modulus have been identified, 
the strongly warped region disappears in the limit of an extremely large 
volume. However, we emphasize and demonstrate quantitatively that, in a 
large intermediate range of the volume modulus, the length of the throat is 
effectively independent of this field. In this parameter range, the universal 
K\"ahler modulus is a brane field which has the peculiar feature of governing 
the size of a very large 4d curvature term localized at the UV brane. 

Our results can be collected in the form of an explicit 5d effective action 
of Randall-Sundrum type, which includes both bulk and brane contributions. 
This action is displayed in Sect.~\ref{ers}, where we also give the relation 
between the main parameters of our 5d formulation and the usual string 
moduli and comment on a possible more complete identification in terms 
of the superfield formulation of the Randall-Sundrum model. 

Section \ref{conc} contains our conclusions and some further comments 
concerning a possible manifestly supersymmetric description in 5 dimensions.

\section{The supergravity solution in the warped region}\label{warpsugra}

To set our notation, we recall that the bosonic sector of type IIB 
supergravity contains the RR and NS 3-forms (we follow the conventions 
of~\cite{Giddings:2001yu,Polchinski:1998rr})
\be
F_3=dC_2 \qquad\mbox{and}\qquad H_3=dB_2
\ee
as well as the RR 5-form
\be
 F_5=dC_4\,.
\ee
These forms enter the leading-order Lagrangian only in the combinations
\bea
\tilde{F}_5&=&F_5-\frac{1}{2}C_2\wedge H_3+\frac{1}{2}B_2\wedge F_3\,,
\label{f5t}\\ \nonumber\\
G_3&=&F_3-\tau H_3\qquad\mbox{where}\qquad \tau=C_0+i\exp(-\phi)\,.
\eea
We will focus on solutions (or regions of solutions) where the RR 0-form
potential $C_0$ vanishes and the dilaton $\phi$ is constant.

The simplest strongly warped region or `throat' available in type IIB 
supergravity is the AdS$_5\times S^5$ geometry arising in the vicinity of a 
stack of $N$ D3 branes (see e.g.~\cite{Verlinde:1999fy,Chan:2000ms, 
Aharony:1999ti} and refs. therein). It can be understood as the deformation 
of flat 10d Minkowski space caused by the D3 branes. More specifically, the 
equations of motion demand~\cite{Schwarz:1983qr,Romans:1984an} that the 
$\tilde{F}_5$ flux on the $S^5$ submanifolds enclosing the branes be 
accompanied by a warp factor $h(r)^{-1/2}$,
\be
h(r)=1+\frac{R^4}{r^4}\,,\qquad\qquad R^4=4\pi g_sN\alpha'^2
\frac{\pi^3}{{\rm Vol}\, S^5}=4\pi g_sN\alpha'^2\,,\qquad \qquad
g_s=e^\phi\,,\label{hads}
\ee
which enters the metric in the form
\be
ds^2=h(r)^{-1/2}dx^2+h(r)^{1/2}(dr^2+r^2d\Omega_5^2)\,.\label{ads5s5}
\ee
Here and in the following, $dx^2$ stands for $\eta_{\mu\nu}\,dx^\mu\,dx^\nu$, 
where $\eta_{\mu\nu}$ is the 4d Minkowski metric. The metric of 
Eq.~(\ref{ads5s5}) is asymptotically flat, but near the branes, i.e. for 
$r\ll R$, becomes the AdS$_5\times S^5$ metric
\be
ds^2=\frac{r^2}{R^2}dx^2+\frac{R^2}{r^2}dr^2+R^2d\Omega_5^2\,.
\ee
It is crucial that the original flat metric (Eq.~(\ref{ads5s5}) with $h(r)$ 
set to unity) is deformed only in a very mild way, which is fully specified 
by the single function $h(r)^{1/2}$. This mildness of the deformation and 
the direct connection between 5-form flux and warp factor persist in 
the much more general geometries considered in~\cite{Giddings:2001yu, 
Giddings:2005ff}. 

A very similar situation arises in the case of the conifold with branes at the 
singularity. The conifold is a real cone over the base $T^{1,1}= (SU(2)\times 
SU(2))\,/\,U(1)$; it has a Calabi-Yau metric described in~\cite{ 
Candelas:1989js}. Topologically, $T^{1,1}\cong S^3\times S^2$, with both 
spheres shrinking to zero size at the singularity at the tip of the cone. Now 
consider the product of 4d Minkowski space with the conifold, such that the 
spacetime metric reads 
\be
\label{conimetric}ds^2=dx^2+dr^2+r^2ds^2_{T^{1,1}}\,.
\ee 
Placing $N$ D3-branes at the conifold singularity corresponds to turning on 
$N$ units of $\tilde F_5$ flux on $T^{1,1}$ in the supergravity background. 
This leads to a deformation analogous to the AdS$_5\times S^5$ case above, 
\be
\label{warpedconimetric}ds^2=h(r)^{-1/2}dx^2+h(r)^{1/2}(dr^2+r^2ds^2_{ 
T^{1,1}})\,,
\ee
with the same function $h(r)$, but with the radius $R$ now given by 
\cite{Herzog:2001xk}
\be
R^4=4\pi g_sN\alpha'^2\frac{\pi^3}{{\rm Vol}\, T^{1,1}}=4\pi g_sN\alpha'^2
\frac{27}{16}\,.\label{hads2}
\ee
For large $r$, far away from the branes, the 
geometry reduces to that of Eq.~\eqref{conimetric}, while near the branes 
it is given by AdS$_5\times T^{1,1}$. 

The AdS/CFT correspondence states that supergravity on this background is dual
to a 4d conformal field theory (for a review of the gauge theory side of the
models we are discussing, see~\cite{Strassler:2005qs}.) 
The radial coordinate $r$ of AdS space 
translates into the renormalization group scale on the gauge theory side, 
with small $r$ corresponding to infrared physics and large $r$ to the 
ultraviolet. To let the throat end in the infrared, i.e. at some small 
but non-zero $r$ and to stabilize the distance to that point, the conformality 
of the above solution in the infrared regime has to be broken. This can be 
realized by turning on $F_3$ flux on the $S^3$ cycle of the $T^{1,1}$. As 
will become clear in a moment, the result will be a radial variation of the 
$\tilde{F}_5$ flux $N$, which is promoted to a function $N_{\rm eff}(r)$. 

One possibility to introduce $F_3$ flux is by placing, in addition to the 
$N$ D3 branes, $M$ fractional D3 branes at the conifold singularity. These 
are actually D5 branes wrapped over the collapsing 2-cycle and are 
constrained to reside at the singular point. They source $M$ units of
flux of the RR 3-form field strength $F_3$ on the $S^3$.

Another possibility to introduce $F_3$ flux is to blow up the conifold 
singularity to give the deformed conifold geometry, in which the $S^3$ 
remains of finite size. Now, $M$ units of $F_3$ flux can be put on the 
$S^3$ cycle. This creates a geometry as in the Klebanov-Strassler 
region~\cite{Klebanov:2000hb} at the end of the throats of~\cite{ 
Klebanov:1999rd,Klebanov:2000nc}. Furthermore, $N_{\rm D3}$ D3 branes 
may be placed in this end-region of the throat.\footnote{
The 
characteristic radius of the Klebanov-Strassler region is given by 
$R^4\sim M^2\alpha'^2g_s^2$. Thus, for $g_sM^2\gg N_{\rm D3}$, the 
Klebanov-Strassler region is relatively flat compared to the curvature caused 
by the $N_{\rm D3}$ D3 branes. One can then interpret this situation as a 
`throat within a throat'~\cite{Cascales:2005rj}, the second throat being a 
pure AdS$_5\times S^5$ region with $N_{\rm D3}$ units of $\tilde{F}_5$ flux. 
However, we will not pursue this more extended geometric picture here.
}

Our main interest is not in the precise way in which the throat is cut off
in the infrared, but in the dependence of $N_{\rm eff}$ on $r$ caused by 
the $F_3$ flux. This dependence, analyzed in detail in~\cite{Klebanov:1999rd, 
Klebanov:2000nc}, is in fact very easy to understand. The key observation is 
that, with $F_3$ flux being present only on the $S^3$ cycle of the $T^{1,1}$, 
\be
(4\pi^2\alpha')^2 N_{\rm eff}(r)=\!\!\!\!\!\int\limits_{T^{1,1}\,\,{\rm 
\scriptstyle at}\,\,r}\!\!\!\!\!\tilde{F}_5 \sim \left(\,\,\int\limits_{S^3\,\,
{\rm \scriptstyle at}\,\,r}F_3\right)\,\left(\,\,\int\limits_{S^2\,\,{\rm 
\scriptstyle at}\,\,r}B_2\right)\,.
\ee
Now, while the $F_3$ flux can not change continuously, the $B_2$ flux will
generically have a non-trivial radial dependence, which is at the origin of 
the the radial variation of $N_{\rm eff}$. 

To be more precise, observe that
\be 
(4\pi^2\alpha')^2(N_{\rm eff}(r_2)-N_{\rm eff}(r_1))=
\!\!\!\!\!\int\limits_{T^{1,1}\,\,{\rm\scriptstyle at}\,\,r_2}\!\!\!\!\!
\tilde{F}_5\,\,-\!\!\!\!\!\int\limits_{T^{1,1}\,\,{\rm\scriptstyle at}\,\,
r_1}\!\!\!\!\!\tilde{F}_5\,\,=\!\!\!\!\!\int\limits_{T^{1,1}\times(r_1,r_2)}
\!\!\!\!\!d\tilde{F}_5\,\,=\!\!\!\!\int\limits_{T^{1,1}\times(r_1,r_2)}
\!\!\!\!\!\!H_3\wedge F_3\,,\label{dneff}
\ee
where the final integrals are over a segment of the throat, which corresponds 
to an interval $(r_1,r_2)$ in terms of the variable $r$. In compactifications 
leading to 4d Minkowski space, $G_3$ is imaginary self-dual~\cite{ 
Giddings:2001yu}. Since the throat will ultimately have to be part of 
such a compactification, we can restrict ourselves to imaginary self-dual 
$G_3$, which implies that $H_3=g_s*_6F_3$. As one can easily see by counting 
the powers of the metric and its inverse, the warp factor drops out of this 
expression for $H_3$. Thus, the last integral in Eq.~\eqref{dneff} can be 
evaluated in terms of $F_3$ and the simple conifold metric: We obtain
\be
\frac{dN_{\rm eff}(r)}{dr}\sim \frac{g_s}{\alpha'^2}
\int\limits_{T^{1,1}\,\,{\rm\scriptstyle at}\,\,r}\!\!\!\!\!d^5y\,
\sqrt{g_{\rm con.}(r,y)}\,F_{mnp}F^{mnp}\,.\label{f32}
\ee
The quantization condition
\be
\frac{1}{2\pi\alpha'}\int_{S^3} F_3=2\pi M
\ee
implies the scaling $F_3\sim M\alpha'$ for the non-vanishing components of 
$F_3$. Taking into account the $r$ dependence introduced by the conifold 
metric and the fact that $F_3$ has no components in the $r$ direction, 
Eq.~(\ref{f32}) evaluates to
\be
\frac{dN_{\rm eff}(r)}{dr}=\frac{ag_sM^2}{r}\qquad\mbox{or}\qquad
N_{\rm eff}(r)=ag_sM^2\ln(r/r_s)\,, \label{nrun}
\ee
where $r_s$ is an integration constant. The ${\cal O}(1)$ numerical prefactor 
 $a=3/(2\pi)$ can be determined using the explicit conifold metric and flux 
forms~\cite{Herzog:2001xk}. Apart from this prefactor, our simplified 
discussion provides the exact result. 

Thus, we have now finally arrived at the geometry of the `conifold throat' 
with varying warp factor. It is characterized by
\be
\label{defconimetric}ds^2=\tilde{h}(r)^{-1/2}dx^2+\tilde{h}(r)^{1/2}(dr^2+
r^2ds^2_{T^{1,1}})\,,
\ee
with 
\be
\tilde{h}(r)=1+\frac{a'\alpha'^2g_s^2M^2\ln(r/r_s)}{r^4}\,,\label{hrun}
\ee
where $a'=4\pi a\cdot 27/16=81/8$. 
This follows by simply inserting $N_{\rm eff}(r)$ in Eq.~(\ref{hads2}). The 
simplicity of this solution is not surprising since it is known that the 
$\tilde{F}_5$ flux, determined by $N_{\rm eff}$, fully specifies the warp
factor~\cite{Giddings:2001yu}. Even though this solution is exact, it is 
not physical. It has a singularity at $r=r_s$, which we assume to be 
resolved by a Klebanov-Strassler region. In other words, we can trust
the above solution for $N_{\rm eff}\gg g_sM^2$ or, equivalently, 
$r\gg r_s$. 

If the Klebanov-Strassler region contains $N_{\rm D3}$ explicit (i.e. not 
simulated by flux) D3 branes, then the throat ends at larger $r$. This is 
clear since the effective D3 charge visible near the end-region is $N_{\rm 
D3}+{\cal O}(g_sM^2)$. Another way to think about this is that, when moving 
along the throat to smaller $r$, one encounters $N_{\rm D3}$ D3 branes. The 
effective charge then suddenly drops by $N_{\rm D3}$ and, subsequently, the 
throat is terminated by a Klebanov-Strassler region `earlier than expected'. 
In summary, if $N_{\rm D3}$ D3 branes are present, we can trust 
Eq.~(\ref{hrun}) only for $N_{\rm eff}\gg N_{\rm D3}+g_sM^2$.\footnote{
Note
that, just from knowledge of $N_{\rm eff}$ and $M$ at large $r$, it is 
impossible to tell whether the throat will end with or without explicit D3 
branes. This is so because contributions to the overall $\tilde{F}_5$ flux 
from $F_5$ (first term on the r.h. side of Eq.~(\ref{f5t})) and from
3-form fluxes (the second and third terms) mix under gauge transformations.
}

In the UV-region, at large $r$, the throat will end when $\tilde{h}(r)$ 
approaches unity, i.e. when $\alpha'^2g_sN_{\rm eff}(r)\simeq r^4$. 
Just from the knowledge of $M$ and $N_{\rm eff}$  at a certain $r$, it 
is impossible to tell where the throat will end in the UV. From the dual CFT 
perspective, this knowledge corresponds to information about 
higher-dimension operators, which is usually hard to access for the 
low-energy observer. In the conical region that follows at larger $r$, 
the integrated $\tilde{F}_5$ flux $N_{\rm eff}$ continues to grow with $r$ 
as before, but the back-reaction is not strong enough to affect the 
geometry. We assume that at some still larger $r=R_c$, the approximate 
conifold geometry goes smoothly over to a compact Calabi-Yau orientifold 
geometry. The compactification radius can thus be approximately identified
with $R_c$. Clearly, the total D3 charge of fluxes and localized sources 
in the bulk of this space has to compensate the $\tilde{F}_5$ flux present 
at the end of the conifold region at $r=R_c$. 

\begin{figure}[ht]
\begin{center}
\includegraphics[width=8cm]{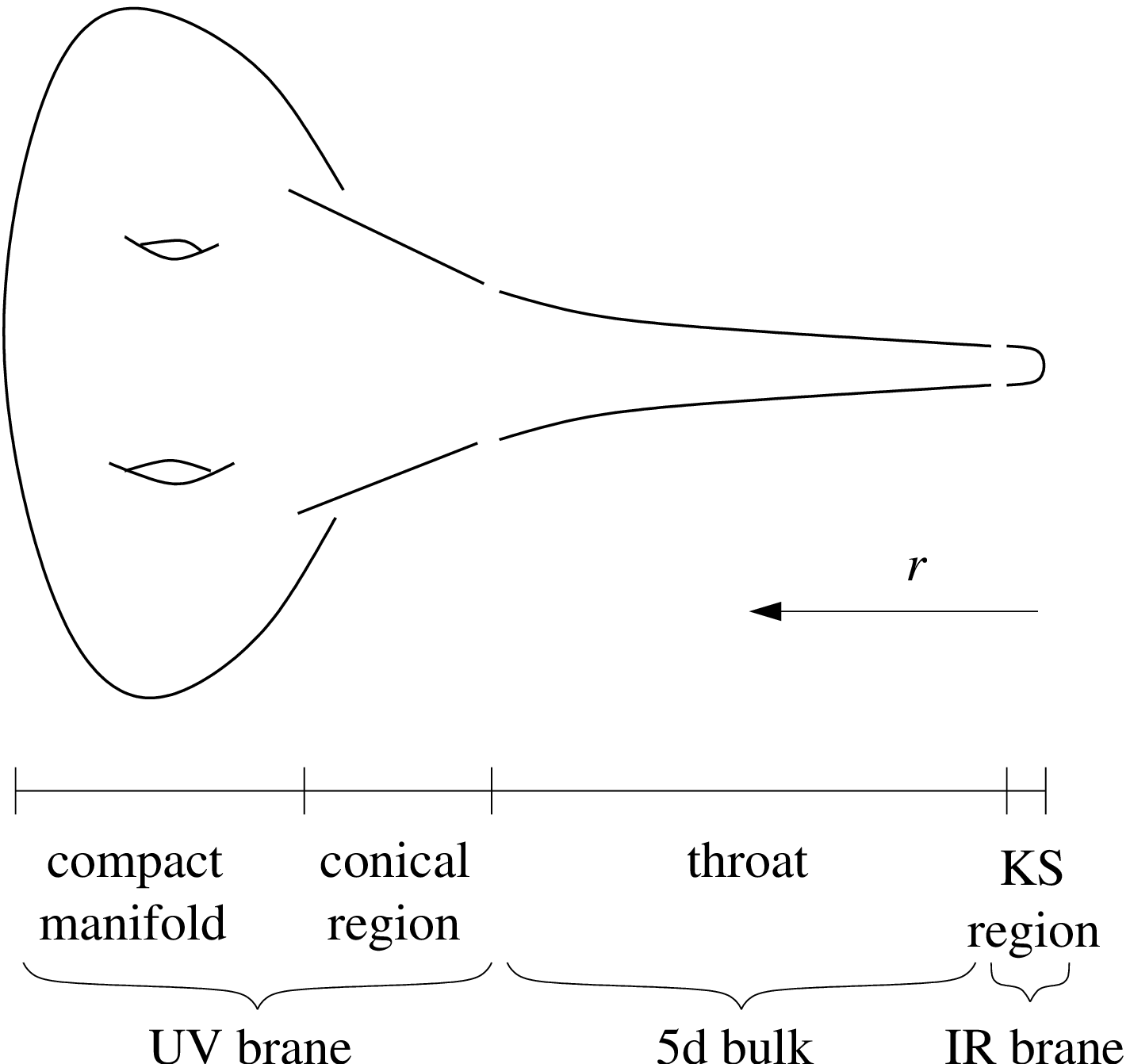}
\end{center}
\refstepcounter{figure}\label{throat}

\vspace*{-.2cm}
{\bf Figure~\ref{throat}:} 
Illustration of the way in which the throat is part of the overall compact 
space. The Klebanov-Strassler (KS) region may contain explicit D3 branes. In
anticipation of the 5d description we have indicated which regions will later
become the UV and IR branes and the 5d bulk.
\end{figure}

We thus arrive at the overall picture illustrated in Fig.~\ref{throat}: The 
compact space (e.g. a Calabi-Yau orientifold) has a conical region (cf. 
Eq.~(\ref{conimetric})) with non-vanishing $\tilde{F}_5$ flux. Going to 
smaller $r$, one reaches the throat region, where the back reaction 
of the flux deforms the geometry significantly and which is finally 
smoothly terminated by a Klebanov-Strassler solution. 

Apart from describing our setting and introducing some notation, the purpose 
of this section was to emphasize the following points for later use: The 
running of $N_{\rm eff}$ with $r$ is driven by the $M$ units of $F_3$ 
and can be derived in a very simple manner, independently of the details of 
the geometry and, in particular, the warp factor. In general, the throat is 
connected to the Calabi-Yau via a conical region, where the running continues 
just as within the throat. The IR end of the throat may contain explicit D3 
branes and may therefore occur at any value of $N_{\rm eff}$ above $N_{\rm 
eff,\,\, min}\sim g_s M^2$.

\section{The equivalent 5-dimensional model}\label{5d}

Consider now the `throat' region of Fig.~\ref{throat} where, over 
sufficiently small distances in the radial direction, the geometry is 
approximately AdS$_5\times T^{1,1}$. This is the region where 
Eqs.~(\ref{defconimetric}) and (\ref{hrun}) apply and where, in the expression 
for the warp factor $\tilde{h}(r)$, the term $\sim 1/r^4$ dominates over the 
additive constant 1. The geometry deviates from AdS$_5\times T^{1,1}$ in that 
the AdS curvature, characterized by the length scale $R_{\rm eff}$, 
\be 
R_{\rm eff}^4(r)=g_sN_{\rm eff}(r)\alpha'^2\frac{27}{16}
=a'\alpha'^2g_s^2M^2\ln(r/r_s)\,,
\ee
varies slowly along the radial direction. The region in question is 
terminated by a Klebanov-Strassler resolution of the conifold in the IR (at 
small $r$) and by the purely conical region together with the compact 
manifold in the UV (at large $r$). 

At a position characterized by $r$, we can give an effective 5-dimensional 
description on length scales $L\gg R_{\rm eff}(r)$. Furthermore, as long as 
$L$ is not too large, 
the change of $R_{\rm eff}$ will be insignificant on length scales 
$L$, so that the 5d geometry will be approximately AdS$_5$. Note, however, 
that there is no length scale at which flat 5d space would provide a good 
approximation. If we want this 5d effective description to be valid in the 
vicinity of the UV brane, we also have to require that the size $R_c$ of 
the compact space (the `brane thickness') is smaller than $L$. Of course, 
this size is governed by the universal K\"ahler modulus, which will be the 
subject of the next section. For the purpose of this section, we simply 
assume that it has been fixed at an appropriate value by some dynamics not 
relevant in the throat region. 

To characterize the throat as a whole, the variation of $R_{\rm eff}$ has to 
be taken into account. In other words, the negative 5d cosmological constant 
$\Lambda_5$, characteristic of AdS$_5$, has to be replaced by a potential 
energy $V(H)$, which must be a function of at least one 5d scalar field $H$ 
to allow for a spatial variation. It is intuitively clear that this field $H$, 
which must have a non-trivial profile $H(r)$ in the bulk, carries the 
information of the quantity $N_{\rm eff}(r)$ or, equivalently, $R_{\rm eff} 
(r)$ of the full 10d picture. As discussed at length in the previous section, 
the microscopic origin of this field is the flux of the NS 2-form potential 
$B_2$ on the $S^2$ cycle of the $T^{1,1}$. 

Working in a 5d Einstein frame with canonically 
normalized $H$,
\be
{\cal L}_5=\frac{1}{2}M_5^3{\cal R}_5-\frac{1}{2}(\partial H)^2-V(H)+\cdots
\,,\label{l5d}
\ee
we can now inquire about the appropriate function $V(H)$. The profile $H(r)$ 
induced by this potential will give rise to a certain scalar-field energy 
density. The back reaction of this energy density has to modify the AdS$_5$ 
geometry in a way reproducing the metric of Eq.~\eqref{defconimetric}. 

To find the potential $V(H)$, it is convenient first to identify an 
alternative radial coordinate which directly measures physical distances 
along the throat. Calling this coordinate $y$, such an infinitesimal distance, 
measured in units of the 5d reduced Planck mass, is given by $M_5\,dy$. To 
compare this with the coordinate $r$, observe that a straightforward 
dimensional reduction of a model with the metric of Eq.~(\ref{defconimetric}) 
to 5d would give rise to an $r$-dependent coefficient of the 5d Ricci scalar, 
which we call $M_{\rm 5,\, eff}^3(r)$. A model with the Lagrangian of 
Eq.~(\ref{l5d}) could only result after a Weyl rescaling by an appropriate 
function of a radially varying scalar field. However, we can avoid this 
procedure by working with the $r$-dependent infinitesimal distance in units 
of $M_{\rm 5,\, eff}(r)$ and demanding
\be 
M_5\,dy=M_{\rm 5,\, eff}(r)\sqrt{g_{rr}}\,dr=[M_{10}^8R_{\rm eff}^5(r)\,
\mbox{Vol}\,T^{1,1}\,]^{1/3}[R_{\rm eff}(r)/r]\,dr\,.
\ee
Using $M_{10}^8=2/[(2\pi)^7\alpha'^4]$ and ${\rm Vol}\,T^{1,1}=16\pi^3/27$, 
this is further evaluated to give 
\be
M_5\,dy=\left(\frac{(g_sM)^4}{3\pi^4}\right)^{1/3}(\ln(r/r_s))^{2/3}\,
d(\ln(r/r_s))\,.
\ee

By the above calculation, $y$ is only determined up to an arbitrary additive 
constant. One possibility of fixing this constant of integration is to define 
\be
M_5\,y=b\left(g_s^2M^2\right)^{2/3}(\ln(r/r_s))^{5/3}=b\left(g_s^2M^2
\right)^{-1}(g_sN_{\rm eff}(r)/a)^{5/3}\,,
\ee
where we have introduced the numerical coefficient $b=3^{2/3}5^{-1}\pi^{-4/3
}$. This is an important intermediate result: We have expressed the basic 5d 
quantity, the physical distance $y$ in units of $M_5$, through $g_s$, $N_{\rm 
eff}(r)$ and $M$, the fundamental dimensionless parameters of the 10d 
geometry. To make the above equations more readable, we define the length 
scale
\be
R_s=M_5^{-1}b\left(g_s^2M^2\right)^{2/3}\,,
\ee
which, up to ${\cal O}(1)$ factors, corresponds to the size of the $T^{1,1}$ 
in the Klebanov-Strassler region. (Of course, our definition remains 
useful if the throat ends at some $r_{IR}\gtrsim r_s$ and no actual 
Klebanov-Strassler region is present.) We can then write
\be
(y/R_s)=(\ln(r/r_s))^{5/3}=\left(\frac{N_{\rm eff}(r)}{ag_sM^2}
\right)^{5/3}\,.\label{ydef}
\ee
In the following, we will treat $y/R_s$ as parametrically large. 

The 5d metric can now be written as
\be 
ds_5^2=e^{2A(y)}dx^2+dy^2\,,\label{5dm}
\ee
where the warp factor, following from Eqs.~(\ref{defconimetric}), 
(\ref{hrun}) and (\ref{ydef}) together with the Weyl rescaling used to go 
to the 5d Einstein frame, reads\footnote{
This 
corresponds to the `special case solution' of Sect.~5 of~\cite{ 
Klebanov:2000nc}, which we therefore find to be the appropriate description 
of the actual throat between the Klebanov-Strassler and the conical 
regions. 
}
\be
A(y)\simeq (y/R_s)^{3/5}+{\cal O}(\ln(y/R_s))+\mbox{const.}
\ee
Here the constant term is irrelevant since it can be absorbed in a rescaling 
of Minkowski space. We may also neglect the subleading logarithmic term, 
writing the warp factor as
\be
\label{rel1} A(y)=k(y)y,\qquad\qquad k(y)\simeq R_s^{-1}(y/R_s
)^{-2/5}\,.
\ee
Note that, up to the sign of $y$ (which is chosen to avoid a large number of
`$-$' signs in the following equations) and the slow variation of $k$, this
corresponds to the now widely used metric conventions 
of~\cite{Randall:1999vf}. 

We are now looking for a potential $V(H)$ such that the back-reaction of 
the varying scalar $H$ induces a varying curvature as in Eq.~\eqref{rel1}. 
In general, such an analysis requires the solution of the coupled equations of 
motion for the metric and $H$. However, in the present case, a simplified 
computation will be sufficient: Because the warp factor varies slowly, we 
can use the equation of motion of a scalar field with potential $V(H)$ in an 
AdS background, allowing for a $y$ dependence of the curvature scale $k$:
\be
\left(\partial_y^2 + 4A'(y)\partial_y \right) H -\frac{\partial V}{\partial H}
=0\,.
\ee
Furthermore, if the typical length scale for the variation of $H$ is larger 
than the curvature radius, we can neglect the second-derivative term:
\be
k(y)\partial_y H\sim \frac{\partial V}{\partial H}\,.\label{heom}
\ee
The value of $k(y)$ is determined by an effective 5D bulk cosmological 
constant coming mainly from the potential term with $H$ set to its local 
VEV. In AdS space, the relation between cosmological constant and curvature 
is~\cite{Randall:1999vf}
\be
V(H)=-24M_5^3 k^2\,,
\ee
which can be solved for $k$ and inserted in Eq.~(\ref{heom}). Using in 
addition the relation $\partial_yH=(dV/dy)(\partial V/\partial H)^{-1}$, we 
arrive at
\be
\frac{\partial V}{\partial H}\sim \left(\frac{-V}{M_5^3}\right)^{1/4}
\left(\frac{dV}{dy}\right)^{1/2}\,.\label{dvdh}
\ee
Since Eq.~(\ref{rel1}) implies that 
\be 
V \sim -M_5^3R_s^{-2}\,(y/R_s)^{-4/5}\,,
\ee
we can rewrite Eq.~(\ref{dvdh}) as 
\be
\frac{\partial V}{\partial H}\sim M^{-21/8}R_s^{3/4}(-V)^{11/8}\,.
\ee

This now finally determines the desired functional dependence of $V$ on $H$:
\be
V(H)\sim -M_5^{7}\,R_s^{-2}\,H^{-8/3}\,. \label{potential}
\ee 
Thus, we conclude that 5d gravity coupled to a scalar field $H$ with the 
potential of Eq.~\eqref{potential} reproduces the effective 5d geometry 
of the throat of the 10D compactification. The $y$-dependence of 
$H$ follows straightforwardly from the above:
\be
H\sim M_5^{3/2}(y/R_s)^{3/10}\,.
\ee
It can also be easily verified that the conditions 
\be 
|\partial_y^2H|\ll |A'(y)\partial_y H|\qquad\mbox{and}\qquad (\partial_y H)^2
\ll |V|\,,
\ee
which justify our simplified treatment, are satisfied. 

\begin{figure}[ht]
\begin{center}
\includegraphics[width=8cm]{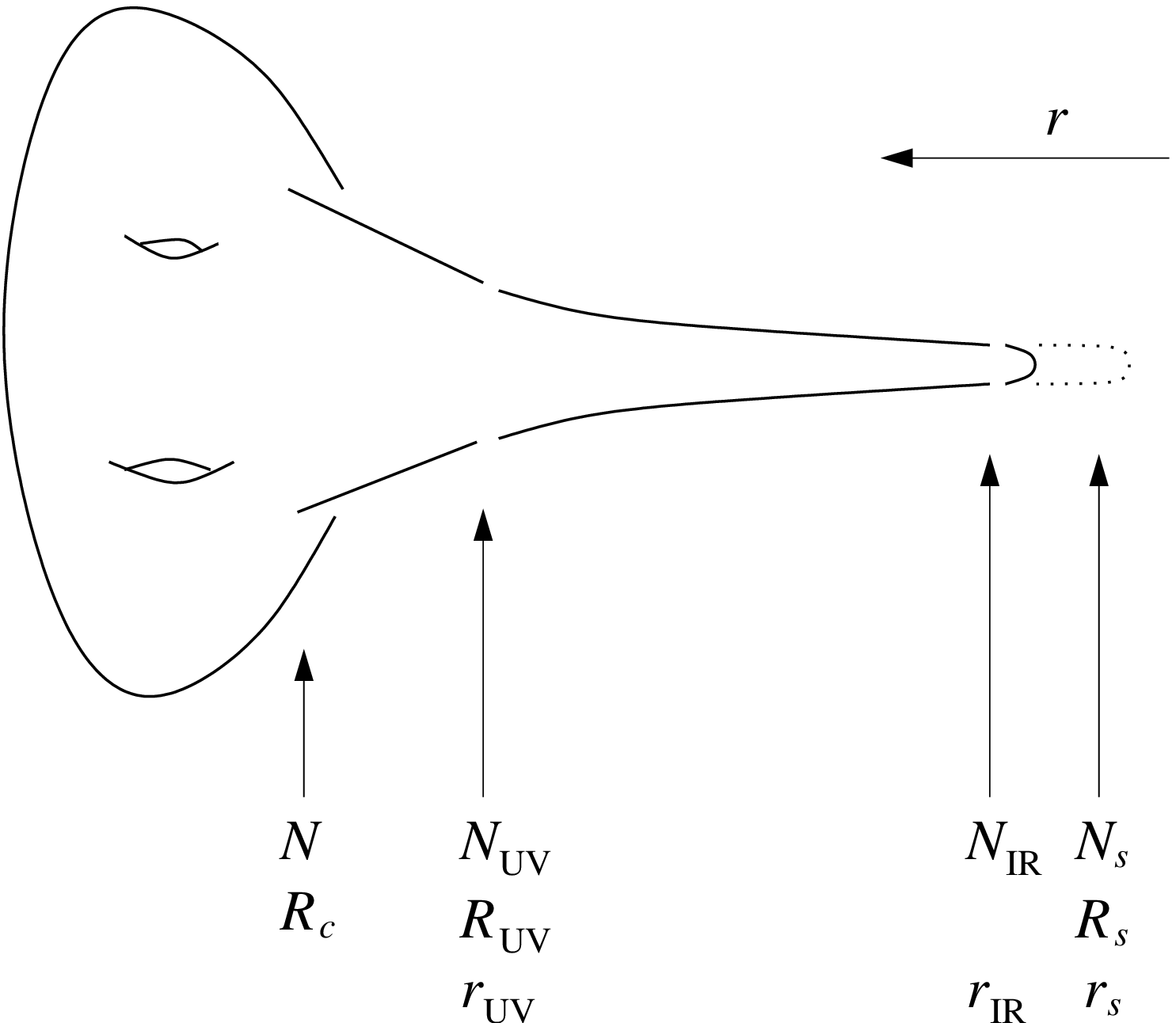}
\end{center}
\refstepcounter{figure}\label{throat2}

\vspace*{-.2cm}
{\bf Figure~\ref{throat2}:} 
The throat with the values of $N_{{\rm eff}}(r)$ and $R_{{\rm eff}}(r)$ at 
several positions $r$. The dotted line indicates that, in the presence of 
explicit D3 branes, the throat may end at $r_{\IR}>r_s$.
\end{figure}

To describe the throat as a whole (cf. Fig.~\ref{throat2}), we need 
to add an IR and UV brane with 
specific tensions and boundary conditions for $H$ to our 5d model. We assume 
the tensions to be positive and negative for the UV and IR brane respectively 
and the values to be such that both branes are static in an AdS space with 
curvature determined by the boundary values of $H$ and $V(H)$. To discuss
the boundary conditions on $H$ explicitly, recall that $H$ substitutes the 
parameters $R_{\rm eff}$ (or, equivalently, $N_{\rm eff}$) of the 10d 
construction. The explicit relations read 
\be 
H\sim M_5^{3/2}(R_{\rm eff}/R_s)^{2}\sim M_5^{3/2}(N_{\rm eff}/N_s)^{1/2}
\qquad\mbox{with}\qquad N_s=ag_sM^2\,.
\ee
Thus, the boundary condition $H(y_s)\sim M_5^{3/2}$ will reproduce the IR end 
corresponding to a Klebanov-Strassler region with $M$ units of $F_3$ flux. 
Field-theoretically, such a boundary condition can be realized by 
an appropriate brane potential for $H$ with an extremely steep minimum. 

Recall that, more generally, the throat may end at $r_{\IR}>r_s$ if explicit 
D3 branes are present at the IR end. The corresponding boundary condition
reads $H(y_{\IR})\sim M_5^{3/2}(N_{\IR}/N_s)^{1/2}$, where $N_{\IR}=N_{\rm 
D3}+ag_sM^2$ can be chosen freely. 

At the UV end, we can define $N$ as the number of $\tilde{F}_5$ flux 
units on the $T^{1,1}$ cycle at the `large' or UV end of the conical region. 
This number is determined by localized sources, e.g. O3 planes and D3 branes,
and regions with 3-form flux within the remainder of the compact 
space.\footnote{
Note
that in the literature $N=MK$ is frequently used to designate the effective 
D3 charge from $M$ units $F_3$ flux on the $S^3$ cycle and $K$ units of $H_3$ 
flux on its dual. This definition coincides with ours if the 3-form flux in 
question is mainly concentrated outside the `compact manifold' of 
Fig.~\ref{throat}.
}
In the conical region, this flux number changes according to Eq.~(\ref{nrun}). 
Assuming that the conical region is not too large, this change is very small
compared to the change that occurs within the throat, so that we can identify 
the $\tilde{F}_5$ flux $N_{\UV}$ at the UV end of the throat (the IR 
end of the conical region) with the flux number $N$ defined above. (A more 
detailed discussion of the subtle effect of the running inside the conical 
region will be given in the next section.) Thus, the UV boundary condition 
of the 5d model reads $H(y_{\UV})\sim M_5^{3/2}(N_{\UV}/N_s)^{1/2}\simeq 
M_5^{3/2}(N/N_s)^{1/2}$. 

In summary, we have presented a 5d model (gravity plus minimally coupled 
scalar field, with the potential given above) which, upon compactification on 
an interval with boundary conditions $H(y_{\IR/\UV})\sim M_5^{3/2}(N_{\IR/
\UV}/N_s )^{1/2}$, provides the 5d description of the conifold throat. In 
particular, the 5d bulk profile of $H$ fixes, together with the boundary 
conditions, the throat length $y_{\UV}-y_{\IR}$. Thus, in the case of the 
conifold throat, flux stabilization corresponds to the familiar 
Goldberger-Wise stabilization mechanism~\cite{Goldberger:1999uk, 
gwback} of 5d Randall-Sundrum models. There is however an important 
difference: in the conifold throat the back-reaction of the scalar field on 
the geometry is crucial. It describes the effect of the $M$ units of $F_3$
flux -- the duality cascade in the dual gauge theory. 

The most important shortcoming of the effective 5d model presented so far is 
the absence of the universal K\"ahler modulus common to such type 
IIB supergravity compactifications~\cite{Giddings:2001yu}. We have avoided 
this issue by simply assuming that the typical radius of the compact space 
at the UV end is somehow stabilized. In the next section, we will relax this 
assumption and discuss the interplay of this degree of freedom with our 5d 
model of the throat.

\section{The universal K\"ahler modulus as a brane-field}\label{kaehlerbrane}

One of the more relevant effects of fluxes in compactifications of type IIB 
supergravity is to generate a potential for complex structure moduli and for 
the dilaton. However, at least one K\"ahler modulus remains massless, even in 
these constructions. In the limit where the warping goes to zero, this modulus 
is simply an overall scaling of the internal metric and hence corresponds to a 
change of the volume of the compact manifold. 

In the presence of warping the scaling behavior is more subtle \cite{ 
Giddings:2005ff}. In terms of the metric of Eq.~\eqref{defconimetric}, the 
flat direction corresponds to a shift $\tilde{h}(r)$ $\rightarrow$ $\tilde{h}
(r)+c-1\,$ for an arbitrary value of the constant $c$. As in the unwarped 
case, this affects the volume of the manifold. 

This realization of the volume modulus can, in fact, be understood very 
easily: The metric of Eq.~\eqref{defconimetric} contains only two 
dimensionful parameters, $\alpha'$ and $r_s$. A volume modulus, if present, 
can only change the ratio of these two scales. Indeed, a rescaling $r_s$ 
$\rightarrow$ $r_sc^{1/4}$ corresponds, together with an appropriate
rescaling of $r$ and $x$, to the shift in $\tilde{h}$ specified above. 

If $c$ becomes extremely large, larger than $\tilde{h}(r_{\IR})$, the throat 
disappears and the variation of $c$ corresponds to an overall scaling of the 
entire compact space. In this regime, the radius $R_c$ is bigger 
than the length scale $L$ at which our 5d effective description is defined. 
In other words, the ``brane thickness'' of the UV brane is so large that the 
5d picture is lost. 

Let us consider instead values of $c$ such that $R_c < L$. In the `compact 
manifold', $\tilde{h}$ is approximately constant and the variation of $c$ 
corresponds again to a simple scaling. As far as the throat is concerned, it 
is crucial that the $\tilde{F}_5$ flux $N$ at the UV end of the conical 
region is fixed; it does not depend on the volume of the compact space. To a 
very good approximation, this is also true for the $\tilde{F}_5$ flux $N_{ 
\UV}$ at the UV end of the throat (i.e. at the IR end of the conical region). 
Furthermore, neither the $F_3$ flux nor the presence of explicit D3 branes at 
the IR end of the throat are affected by the volume scaling. Thus, neither the 
boundary condition $N_{\UV}$ nor $N_{\IR}$ changes when $c$ varies 
and then, as we discussed in Sect.~\ref{5d}, the length of the throat is 
fixed. This means that, in the 5d description, the K\"ahler modulus plays the 
role of a massless UV-brane field while the 5d radion is already stabilized. 

However, this picture is correct only at first approximation. The key to the 
$c$-independence of $N$ was its definition as the flux at the transition 
point between the conical and the more general compact geometries. This 
definition does not depend on the overall scaling. By contrast, $N_{\rm UV}$ 
is defined at the transition point between conical and throat geometries. 
As we will now demonstrate, this transition point has non-trivial 
$c$-dependence, which is reflected in a weak $c$-dependence of $N_{\rm UV}$. 
The resulting effect on the length of the throat is small compared to the 
effect on the compact region, as we will show explicitly. Nevertheless, 
for extremely large $c$ this effect will cut into the length of the throat 
such that, eventually, the throat disappears. This is consistent with the 
limit of weak warping discussed above. 

We refer to the combination of compact space and conical region 
as the UV brane. From the 10d point of view, this is the area 
where the warp factor is, to a good approximation, constant. Thus, the 
universal K\"ahler modulus simply corresponds to an overall rescaling of 
this space. In particular, the flux number
\be
N=N_{\rm eff}(R_c)\simeq\frac{1}{(4\pi^2\alpha')^2}\int\limits_{T^{1,1}\,\,{ 
\rm at} \,\,r=R_c}\!\!\!\!\tilde{F}_5
\ee
is invariant under this rescaling. From the point of view of the throat, it 
is determined, once and for all, by the localized sources and flux in the 
compact space, i.e. `on the other side' of the $T^{1,1}$ submanifold at 
$r=R_c$. 

We now focus on the conical region and the throat. If we insist on always 
using coordinates such that the warp factor in the conical region is unity, 
as in Eqs.~(\ref{defconimetric}) and (\ref{hrun}), then $R_c$ can be 
identified with the universal K\"ahler modulus. Given the general 
$r$-dependence of $N_{\rm eff}$ in throat and conical region, 
Eq.~(\ref{nrun}), one finds the constraint 
\be
N=ag_sM^2\ln(R_c/r_s)\,,
\ee
which fixes $r_s$ in terms of $R_c$. This gives the warp factor (cf. 
Eq.~(\ref{hrun}))
\be
\tilde{h}(r)=1+\frac{(a'/a)\alpha'^2g_s[N-a g_sM^2\ln(R_c/r)]}{r^4}\,.
\label{hrun1}
\ee
The boundary between conical region and throat, $r=r_{\UV}$, is then 
determined by the solution of the equation $\tilde{h}(r_{\UV})=1$. Assuming 
that, at this boundary, the logarithmic term in Eq.~(\ref{hrun1}) is small 
relative to $N$ and working to leading order in this small term, we 
find 
\be
r_{\UV}^4\simeq (a'/a)\alpha'^2g_s\left[N-\frac{a g_sM^2}{4} 
\ln\left(\frac{R_c^4}{(a'/a)\alpha'^2g_sN}\right)\right]\,.\label{ruv}
\ee
Thus, the conical region shrinks to zero size if $R_c^4$ takes the value 
\be
R_{c,\,{\rm min}}^4=(a'/a)\alpha'^2g_sN\,,
\ee
and our approximation remains valid as long as $(R_c/R_{c,\,{\rm min}})\ll
\exp(N/a g_sM^2)$. In the large-hierarchy case, which is our main 
interest in this paper, the last expression is roughly equal to the inverse 
warp factor and is thus very large. In other words, there is a large range in 
which the variation of $R_c$ (i.e. of the universal K\"ahler modulus) 
has very little effect on the throat length, as expressed by Eq.~(\ref{ruv}). 
In this domain, it is mainly just a scaling of the compact manifold at the UV 
end of the throat. Thus, we are led to the conclusion that, from the 5d 
point of view, the universal K\"ahler modulus is a field localized at the UV 
brane.

Let us now translate the above discussion to the 5d picture in a more 
quantitative way. From the 5d perspective, the fundamental scale is the 
reduced 5d Planck mass defined by $M_5$. Near the UV brane, $M_5$ is related 
to $M_{10}$ by $M_5^3\simeq M_{10}^8R_{\UV}^5$. For not too large $R_c$, we 
can identify $R_{\UV}$ with $R_{c,\,{\rm min}}$, with the result that 
\be
R_cM_5\sim (g_sN)^{2/3}(R_c/R_{c,\,{\rm min}})\,.\label{tth}
\ee
We can think of this as of the UV brane thickness in units of $M_5$. In the 
same units, the physical length $L_{\rm throat}$ of the throat is given by 
\be
\label{lth} 
L_{\rm throat}M_5\sim(y_{\UV}-y_{\IR})M_5\sim (g_sM)^{4/3}\left[\left(\frac{ 
N_{\UV}}{ag_sM^2}\right)^{5/3}-\left(\frac{N_{\IR}}{ag_sM^2}\right)^{5/3}
\right]\,,
\ee
where $N_{\UV}$ is the flux at the small end of the conical region or, 
equivalently, at the large end of the throat. Our interest is in the 
dependence of $L_{\rm throat}$ on $R_c$. Thus, we cannot simply identify 
$N_{\UV}$ with $N$, but rather we have to take care of this subtle distinction 
which is due to the running in the conical region:
\be
N=N_{\UV}+ag_sM^2\ln(R_c/R_{c,\,{\rm min}})\,.
\ee
We now assume that $R_c$ grows by a factor $1+\epsilon$ (where $\epsilon\ll 
1)$. Then, on the one hand, the thickness of the UV brane in units of $M_5$ 
increases by $\sim\epsilon(g_sN)^{2/3}R_c/ R_{c,{\rm min}}$.
 On the other hand, the length of the throat, also measured in units of 
$M_5$, shrinks by $\sim\epsilon (g_sM)^{4/3}(N_{\UV}/ag_sM^2)^{2/3}$. The 
ratio of these two quantities is $\sim R_c/ R_{c,{\rm min}} >1$, i.e. the 
throat shrinks less than the brane thickness grows. 

This can be turned into an even more explicit argument for the K\"ahler 
modulus being a brane field: From the 5d perspective, it is perfectly 
acceptable to define the throat length either by Eq.~(\ref{lth}) or, 
including the UV brane thickness into the size of the 5d interval, by 
the sum of Eqs.~(\ref{tth}) and (\ref{lth}). When $R_c$ grows, the throat 
length shrinks according to the first and grows according to the second 
definition. Thus $R_c$ cannot be consistently identified
with the length of the 5d interval. Instead, it
has to be modelled by a field localized at the UV brane. 
Of course, because our 5d effective theory is valid only at length scales 
above $L$, we should be careful not to increase $R_c$ above $L$. Otherwise, 
the 5d description of the UV end becomes meaningless.

\section{The explicit Randall-Sundrum-type 5d action}\label{ers}

We are now finally in a position to construct the 5d effective action 
including bulk and brane fields. For $R_c\gg R_{c, {\rm min}} \simeq R_{\UV}$, 
the integral over the compact space at the UV end of the throat contributes 
\be
{\cal L}_{\UV}=\frac{1}{2}M_{10}^8R_c^6\left({\cal R}_4+30(\partial \ln R_c)^2
+\cdots\right)\label{luv}
\ee
to the 4d effective action.\footnote{
The 
prefactor 30 arises as $k(k-1)$, with $k=6$ the number of compact dimensions.
}
We can view this as a precise definition of $R_c$, which is chosen such that 
$R_c^6$ is the volume of the compact space. Note that, at this point, the 4d 
metric implicit in ${\cal R}_4$ is simply the 4d part of the 10d metric in 
the Einstein frame, i.e. no Weyl rescaling has been performed. 

The bulk part was already given in Eq.~(\ref{l5d}). Writing the 5d metric as 
\be
ds_5^2=e^{2A(y)-2A(y_{\UV})}g_{\mu\nu}dx^\mu\,dx^\nu+dy^2\,,\label{5d4d}
\ee
and integrating from $y_{\IR}$ to $y_{\UV}$, this contributes the following 
piece to the Einstein-Hilbert-term of the 4d action: 
\be 
\frac{1}{2}\left(M_5^3\int_{y_{\IR}}^{y_{\UV}}dy\,\exp\left[2\left(\frac{y}
{R_s}\right)^{3/5}\!\!-2\left(\frac{y_{\UV}}{R_s}\right)^{3/5}\right]
\right)\,{\cal R}_4\approx\frac{5}{12}M_5^3R_s\left(\frac{y_{\UV}}{R_s}
\right)^{2/5}{\cal R}_4\,.\label{5dint}
\ee
Here ${\cal R}_4$ is to be evaluated with the 4d metric $g_{\mu\nu}$. The 
warp factor in Eq.~(\ref{5d4d}) has been normalized to ensure consistency 
with the 4d metric in Eq.~(\ref{luv}).

Note that the relative normalization of the coefficients of the ${\cal R}_4$ 
and the $(\partial\ln R_c)^2$ terms in Eq.~(\ref{luv}) is intimately linked 
to the fact that $R_c^4$ is the imaginary part of a superfield $\rho$~\cite{ 
Giddings:2001yu}, which is part of a no-scale supergravity model~\cite{ns}. 
This relation is apparently destroyed by the addition of the 4d 
Einstein-Hilbert contribution of Eq.~(\ref{5dint}). However, this term is 
subdominant in the large-$R_c$ limit in which Eq.~(\ref{luv}) was derived. 
Corrections to this equation are indeed expected since, near the IR end of 
the conical region, $R_c$ loses its interpretation as an overall scaling 
modulus of the compact space. For consistency with Eq.~(\ref{5dint}), the 
coefficient of the ${\cal R}_4$ term in  Eq.~(\ref{luv}) has to be modified 
according to
\be 
M_{10}^8R_c^6 \to M_{10}^8R_c^6-\frac{5}{6}M_5^3R_s\left(\frac{y_{\UV}}{R_s}
\right)^{2/5}\,.
\ee

After these remarks we now give the full 5d action to the extent that it can 
be inferred from the present analysis. In doing so, it is convenient to 
absorb a factor $g_sM$ into the definition of the scalar field. Thus, we 
define
\be 
\tilde{H}=cg_sMH\,,
\ee
where the ${\cal O}(1)$ numerical constant $c$ is chosen to ensure that 
$\tilde{H}(y)=M_5^{3/2}(g_sN_{\rm eff}(y))^{1/2}$ for the solution of 
Sect.~\ref{5d}. The action now reads
\bea
S_{5d}&=&\int d^5x\sqrt{-g_5}\left(\frac{1}{2}M_5^3{\cal R}_5-\frac{1}{2
(cg_sM)^2}(\partial\tilde{H})^2+c_VM_5^9\tilde{H}^{-8/3}+\cdots\right)
\label{ba}\\
&&+\int\limits_{\rm UV\,\,brane}\!\!\!\!d^4x\sqrt{-g_{4,\,\UV}}
\left(K_{\UV}+{\cal L}_{\UV}\right)
+\int\limits_{\rm IR\,\,brane}\!\!\!\!d^4x\sqrt{-g_{4,\,\IR}}
\left(K_{\IR}+{\cal L}_{\IR}\right)\,,\nonumber 
\eea
where $K_{\UV/\IR}$ is the trace of the extrinsic curvature (the 
Gibbons-Hawking surface term~\cite{Gibbons:1976ue}) and $(g_{4,\,\UV 
/\IR})_{\mu\nu}$ is the induced metric at each of the 4d boundaries. The 
value of the positive ${\cal O}(1)$ numerical constant $c_V$ is not important 
for our purposes. The brane Lagrangians are
\bea
{\cal L}_{\UV}&=&\frac{c_1}{2}M_5^2(g_sN_{\UV})^{-10/3}\left[\left(
(R_cM_5)^6-c_2(g_sN_{\UV})^4\right){\cal R}_4+30(R_cM_5)^6(\partial 
\ln R_c)^2\right]
\nonumber \\
&&-V_{\UV}(\tilde{H})-\Lambda_{4,\,\UV}+\cdots
\eea
and
\vspace*{.2cm}
\be
\hspace*{-8.5cm}{\cal L}_{\IR}\,\,\,\,=\,\,\,\,\,-V_{\IR}(\tilde{H})-
\Lambda_{4,\,\IR}+\cdots\,,
\ee
with numerical coefficients $c_1=32\pi^{1/3}/9$ and $c_2=3\cdot 2^{2/3}/(32
\pi)$. Here $V_{\UV}$ and $V_{\IR}$ are steep potentials setting $\tilde{H}$ 
to its values at the UV and IR brane respectively, for example,
\be
V_{\UV/\IR}=\mu^2\left[\tilde{H}-M_5^{3/2}(g_sN_{\UV/\IR})^{1/2}\right]^2\,,
\ee
with a very large coefficient $\mu$. The brane tensions or 4d brane 
cosmological constants $\Lambda_{\UV}$ and $\Lambda_{\IR}$ have values
\be
\Lambda_{\UV}=+M_5^4\sqrt{6/c_V}(g_sN_{\UV})^{-2/3}\qquad\mbox{and}\qquad
\Lambda_{\IR}=-M_5^4\sqrt{6/c_V}(g_sN_{\IR})^{-2/3}\,.
\ee

The fundamental dynamics of the throat can now be easily understood from the 
5d action of Eq.~(\ref{ba}): The scalar field $\tilde{H}$ governs, via the 
potential term, the (approximately AdS) curvature and hence the warping. The 
rapidity with which the curvature changes as one moves along the 5th 
dimension is determined by the coefficient of the kinetic term for $\tilde{H}
$. In the limit of vanishing $M$, no change is possible -- this is the pure 
AdS$_5$ case. The boundary or brane values of $\tilde{H}$ are determined 
by steep brane potentials. The IR-brane potential models the way in which 
the Klebanov-Strassler region (or a more complicated corresponding geometry) 
determines the value of $N_{\rm eff}$ in the IR regime. The UV-brane potential 
models the way in which the various stringy and field-theoretic sources of 
D3-brane flux in the compact space determine $N_{\rm eff}$ in the conical 
region. The combined dynamics of UV/IR-brane and 5d bulk actions then 
stabilizes the length of the interval and fixes the hierarchy. 

In the above 5d effective action, $R_c$ appears as a brane field localized at 
the UV-boundary. However, it is a brane field of very peculiar type. In the 
5d Einstein frame, $R_c$ is part of the coefficient of the brane-localized 
Ricci-scalar and has a wrong-sign kinetic term. Of course, this can 
be remedied by performing an appropriate $R_c$-dependent Weyl rescaling of 
the 5d metric. However, in such a Weyl frame $R_c$ would cease to be a 
UV-brane field. Note furthermore that $R_c$ can easily be parametrically 
larger than its lower bound (in the present analysis) $R_{c,\,{\rm min}} 
\simeq R_{\UV}$. In this case, our 5d model develops a parametrically 
large gravitational brane-kinetic term. This interesting 
possibility~\cite{bkin} has been considered for phenomenological reasons in 
field-theoretic model building (see e.g.~\cite{bkinpheno}). 

Before closing, we would like to explicitly relate the most important 
parameters of our 5d description, the boundary scalar $R_c$ and the 5d 
radion $\Delta y=y_{\UV}-y_{\IR}$, to the corresponding standard string 
moduli. Focussing on the universal K\"ahler modulus $\rho$ and a single 
complex structure modulus $z$ (and neglecting the warping for the moment),
the 4d ${\cal N}=1$ superfield action is determined by the K\"ahler 
potential
\be 
{\cal K}(\rho,z)=-3\ln[-i(\rho-\bar{\rho})]-\ln\left(-i\int\Omega
\wedge\bar{\Omega}\right)\,,\label{kal}
\ee
and the superpotential 
\be
W(z)=\int G_3\wedge\Omega
\,.
\ee
The holomorphic (3,0) form $\Omega$ is normalized using some 3-cycle of the 
compact space at the UV end of the throat, and $z$ is defined via the $S^3$ 
cycle in the throat discussed in Sect.~\ref{warpsugra},
\be
z=\int_{S^3}\Omega\,.\label{zperiod}
\ee

It is well-known that the imaginary part of the universal K\"ahler modulus 
governs the compactification volume. More precisely, the 4d no-scale field 
$\rho$ of~\cite{Giddings:2001yu} is related to $R_c$ by
\be
\Im\rho\sim R_c^4\,.
\ee
We can leave the constant of proportionality arbitrary since we do not 
intend to fix a possible additive constant in ${\cal K}$. 

In~\cite{Giddings:2001yu} the relation of the complex structure modulus $z$ 
to the relative warping between the UV and IR region is found to
be
\be
e^{A(r_{IR})-A(r_{\UV})} \simeq |z|^{1/3}\,.\label{aaz}
\ee
Here $\exp[2A(r)]=\tilde{h}(r)^{-1/2}$ (cf. Eq.~(\ref{defconimetric})) is the 
10d warp factor, which differs from the 5d warp factor $\exp[2A(y)]$ of 
Eq.~(\ref{5dm}) by an insignificant (non-exponential) correction related to 
the 5d Weyl rescaling. The relative 5d warping is 
\be
e^{A(y_{IR})-A(y_{\UV})} \simeq \exp\left[-(\Delta y/R_s)^{3/5}\right]\,,
\label{aad}
\ee
which allows us to express $z$ through the 5d radion:
\be
|z|^{1/3}\simeq \exp\left[-(M_5\,\Delta y/b)^{3/5}(g_sM)^{-4/5}\right]\,.
\ee
This concludes our comparative discussion of $R_c$ and $\Delta y$ and the 
string moduli $\rho$ and $z$. 

Of course, it would be desirable not to stop here but rather to go on and 
identify the superfield description of the stabilized Randall-Sundrum 
model~\cite{Luty:2000ec} with the moduli description of the 10d flux 
compactification. At present, we can only offer some comments which may lead 
in this direction:

The essential quantity on the 5d side is the radion superfield $T$ with 
Re$\,T\sim \Delta y$. The K\"ahler potential in terms of $T$ is expected to 
be~\cite{Lalak:2001dv} (see also~\cite{Luty:2000ec,Marti:2001iw})
\be 
{\cal K}_{5d}\simeq -3\ln\left[\int^{y_{\UV}}_{y_{\UV}-{\Re T}}dy\,e^{2A(y)-2
A(y_{\UV})}\right]\,,
\label{lint}
\ee
i.e. it is proportional to the logarithm of the coefficient of the Ricci 
scalar in the 4d effective action before Weyl rescaling. We now consider 
$y_{\UV}$ to be constant and focus exclusively on the $T$ dependence 
entering through the lower integration limit $y_{\IR}=y_{\UV}-\Re T$. This 
$T$ dependence corresponds to the $z$ dependence in the language of 10d 
moduli (cf. Eqs.~(\ref{aaz}) and (\ref{aad})) so that we can write 
\be
\int\limits^{y_{\UV}}_{y_{\UV}-{\Re T}}\!\!\!\!\!dy\,e^{2A(y)-2A(y_{\UV})}
\,\,\,=\,\,\,{\rm const.}-|z|^{2/3}\!\int\limits_{-\infty}^{y_{\IR}}\!
dy\,e^{2A(y)-2A(y_{\IR})}\,\,\,\simeq\,\,\,{\rm const.}-\frac{|z|^{2/3}}
{2A'(y_{\IR})}\,.
\ee
Since $A'(y_{\IR})\sim(-\ln|z|)^{-2/3}$, this implies for the $z$-dependent
part of the K\"ahler potential
\be\label{5dkp}
{\cal K}_{\rm 5d}\simeq-3\ln\left[{\rm const.}\,-\,|z|^{2/3}
(-\ln|z|)^{2/3}\right]\sim |z|^{2/3}(-\ln|z|)^{2/3}\,,
\ee
where the prefactor and subdominant terms have been suppressed.

This is to be compared with the $z$-dependent part of Eq.~(\ref{kal}) which,  
following~\cite{Frey:2003dm}, can be computed as follows: To account for 
warping, $\Omega\wedge\bar{\Omega}$ is replaced with $e^{-4A}\Omega 
\wedge\bar{\Omega}$~\cite{DeWolfe:2002nn}. The dominant $z$-dependent 
contribution comes from the tip of the throat and depends only on two 
period integrals. The relevant cycles of the compactification manifold are 
the conifold 3-cycle with period $z$, cf. Eq.~\eqref{zperiod}, and its dual 
$\tilde S^3$ with period
\be
\int_{\tilde S^3}\Omega=\frac{z}{2\pi i}\ln z+{\rm holomorphic}
\ee
($\tilde S^3$ will extend outside the throat into the compact manifold, whose 
precise form determines the holomorphic part). There will in general be
other pairs of 3-cycles with period integrals that depend purely 
holomorphically on $z$. With the warp factor contribution at the tip given by
$e^{-4A}\sim |z|^{-4/3}$, we obtain for the $z$-dependent part
\be\label{harm}
-\ln\left(-i\int\,e^{-4A}\,\Omega\wedge\bar\Omega\right)\simeq\ln\left[{\rm const.}-
|z|^{2/3}\ln(z\bar{z})+\cdots\right]\sim |z|^{2/3}(-\ln|z|)\,.
\ee
Here the ellipses stand for higher-order terms of the form $f(z)\bar g(\bar z)$ 
with $f,g$ holomorphic. As before, the prefactor and subdominant terms have 
been suppressed.

We see that, in the small-$z$-limit, the structure of Eqs.~(\ref{5dkp}) and 
(\ref{harm}) agrees (in the sense that the logarithms of the derivatives of 
$\cal K$ coincide). Going beyond this approximation (which on the 5d side 
corresponds to constant warping), the two results are still intriguingly 
close but not quite the same. The failure to fully match the string-theoretic 
with the 5d field-theoretic result is not unexpected in many ways. On the one 
hand, it may be necessary to account for subleading warping corrections on 
the 10d side. On the other hand, calculating the K\"ahler potential on the 
basis of Eq.~(\ref{lint}) and using the naive identification of $\Delta y$ 
in terms of $|z|$ may be too simplistic. It may be necessary and it would 
certainly be highly desirable to start with a manifestly supersymmetric 5d 
Lagrangian which reproduces the correct 5d scalar potential governing the 
profile of the Goldberger-Wise scalar $H$ and hence the warp factor. We leave 
the detailed analysis of these issues to future work.

\section{Conclusions and Outlook}\label{conc}

In this paper, we have derived what we believe to be the main characteristics 
of the 5d effective theory describing the throat region of a type IIB flux 
compactification. It is well known that, at first approximation, the throat 
can be viewed as a Randall-Sundrum I model, where the UV brane is the compact 
space and the IR brane is the Klebanov-Strassler region of the throat. We 
take this analogy beyond leading order by identifying the 5d dynamics that 
leads to the deviation from the AdS$_5$ geometry observed in flux 
compactifications and to the stabilization of the size of the 5th dimension, 
i.e. the radion.

To be more specific, we find that in the effective 5d theory the radion is 
stabilized by a variant of the Goldberger-Wise mechanism: The 5d bulk scalar 
$H$ has a potential $V(-H)\sim H^{-8/3}$ which induces a non-trivial 
5d bulk profile of $H$. Via gravitational back-reaction, this gives rise to 
a 5d curvature consistent with the known 10d throat solution. The non-trivial 
profile of $H$ reflects the variation of the size of the $T^{1,1}$ transverse 
space as one moves along the throat (or, equivalently, of the 5-form flux on 
the $T^{1,1}$, or of the NS 2-form flux on the $S^2$ cycle of the $T^{1,1}$). 
Together with the UV and IR boundary values of $H$, which are fixed by flux 
numbers and (anti-) D3 brane charges, this profile determines the length of 
the throat.

From the 5d perspective, the universal K\"ahler modulus (which can be left 
unfixed for our purposes) is a UV brane field. It governs the coefficient 
of a UV-brane-localized 4d Ricci-scalar~\cite{bkin}. As the K\"ahler modulus 
grows, a very large brane-localized gravitational kinetic term develops, 
which might have interesting phenomenological and cosmological 
implications~\cite{bkinpheno}. At the same
time, the effective brane thickness of the UV brane grows. It is then clear 
that, for extremely large volumes, the UV brane `eats up the throat' and 
both the hierarchy and the 5d picture are lost. However, in a large 
intermediate range of volumes, the length of the throat is practically 
independent of the universal K\"ahler modulus. 

Clearly, our analysis leaves many questions unanswered. First of all, it 
would certainly be desirable to derive the 5d action by an explicit 
dimensional reduction rather than by consistency arguments, as we have done. 
This would, in particular, enable us to include the full set of light 5d 
fields in the 5d Lagrangian and to characterize the dynamics of the UV and 
IR brane in more detail. More importantly, such an explicit calculation may 
open the way to a better understanding of the supersymmetry that should 
be a feature of our 5d bulk action. 

More specifically, we are faced with the following problems as far as 5d 
supersymmetry is concerned. The conifold throat with constant warping is 
known to have 4d ${\cal N}=1$ superconformal symmetry~\cite{Klebanov:1998hh, 
Romans:1984an}, which is referred to as ${\cal N}=2$ (or 5d ${\cal N}=1$) 
SUSY in the literature concerned with the supersymmetric Randall-Sundrum 
model~\cite{srs}. Since the potentials in such a theory are highly 
constrained, it should be non-trivial and interesting to understand how 
our effective $H^{-8/3}$ potential can arise. Unfortunately, the recently 
discussed supersymmetric Goldberger-Wise models~\cite{Maru:2003mq} based on 
massive bulk hypermultiplets do not appear to generate such a potential in 
any obvious way. In fact, one might consider the alternative possibility 
that, because of the $M$ 3-form flux units on the $S^3\subset T^{1,1}$, SUSY 
is {\it always} broken from the 5d point of view. In this case, there would 
be no 5d supersymmetric Lagrangian. However, it is then unclear how the 
effective 4d (non-conformal) ${\cal N}=1$ SUSY, which is known to be present 
in the 4d effective theory, arises. We consider these to be interesting and 
important problems for the future.

A better understanding of our stepwise (10d to 5d to 4d) dimensional 
reduction in a manifestly supersymmetric approach may be relevant for the 
analysis of SUSY breaking mediation in this framework. Given the large 
expectations that have been placed on geometric or conformal sequestering 
and the corresponding interest in its possible violation (see 
e.g.~\cite{Luty:2000ec,seq}), we consider it important to confront those 
models with the explicit string-theory realization of the Randall-Sundrum 
scenario discussed here. We hope that new insights in 5d and 10d SUSY breaking 
phenomenology will be possible along the lines of the 5d effective field 
theory approach discussed in this paper. 

\noindent
{\bf Acknowledgements}:\hspace*{.5cm}We would like to thank J.~Erdmenger, 
J.~March-Russell, 
\linebreak
H.-P.~Nilles, T.~Noguchi, M.~Olechowski, R.~Rattazzi, C.~Scrucca, 
S.~Stieberger, A.~Uranga, A.~Westphal, and A.~Zaffaroni for helpful 
discussions. 


\end{document}